\documentclass[twocolumn,secnumarabic,amssymb, amsmath]{revtex4}

\usepackage{graphicx}
\usepackage{color}
\usepackage{axodraw}
\usepackage{psfrag}
\usepackage{amssymb}

\def\Amix{{A}_{\rm mix}}
\def\Adir{{A}_{\rm dir}}
\def\Adel{{A}_{\Delta\Gamma}}

\newcommand{\sss}{\scriptscriptstyle}
\newcommand{\dis}{\displaystyle}
\newcommand{\eq}[1]{\begin{equation} #1 \end{equation}}
\newcommand{\eqa}[1]{\begin{eqnarray} #1 \end{eqnarray}}

\newcommand{\sect}[1]{\section{\hspace{-0.3cm} #1}}

\begin{document}

\title{An analysis of $B_{d,s}$ mixing angles in presence of New Physics\\ and an update of $B_s \to \bar{K}^{0*} K^{0*}$}

\author{S\'ebastien Descotes-Genon$^{a}$, Joaquim Matias$^{b}$ and Javier
Virto$^{b}$}
\affiliation{$^{a}$ Laboratoire de Physique
Th\'eorique, CNRS/Univ. Paris-Sud 11 (UMR 8627),
91405 Orsay Cedex, France \\
$^{b}$ Universitat Aut\`onoma de Barcelona,
08193 Bellaterra, Barcelona, Spain}

\begin{abstract}
We discuss a simple approach to measure the weak mixing angles $\phi_s$ and $\phi_d$ of the 
$B_s$ and $B_d$  systems in the presence of New Physics. We present a 
new expression that allows one to measure directly $\phi_{d,s}^{\rm NP}$
 if New Physics contributes significantly to the mixing only. We apply the method to specific penguin-mediated $B\to PP$, $B\to PV$ and $B \to VV$ modes. We provide a very stringent and simple bound on the direct CP asymmetries of all these modes, the violation of which is a signal of New Physics in decay.
Within the same theoretical framework,  an updated prediction for the branching ratio of $B_s \to K^{0*} {\bar K} ^{0*}$ is presented, which can be compared with a recent LHCb analysis.
\end{abstract}


\maketitle

In our quest for New Physics (NP) in flavour processes, the weak angles $\phi_M$ involved in the mixing of $B_d$ ($\phi_d$) and $B_s$  ($\phi_s$) mesons have been studied with
a great attention recently, since they have shown interesting discrepancies with respect to the Standard Model~\cite{CKM,Charles:2011va}. On the one hand, there is a tension in the SM fit between the measurement of $\phi_d$ and the branching ratio of  $B\to \tau \nu$
(see also Refs.~\cite{lunghisoni,gino} where the discrepancies with respect to other inputs are discussed, e.g., inclusive or exclusive semileptonic decays, or mixing-dependent observables). On the other hand, the most recent angular analyses of $B_s\to J/\Psi\phi$ by CDF, D{\O} and LHCb \cite{CDF,D0,LHCb} seem to reduce substantially the space for New Physics in $\phi_s$. Finally, there still remains an important discrepancy with the Standard Model through the dimuon asymmetry measured at D{\O}~\cite{D01,D02}.

In this context, it is useful to devise alternative methods to extract the mixing angle of both $B_d$ ($\phi_d$) and $B_s$  ($\phi_s$) systems from experiment.
In this paper we will take the point of view of assuming that New Physics contributes significantly only to the mixing phases whereas the formalism of the CKM matrix can still be used to analyse decay amplitudes. Under this assumption we present a clean method to extract the NP contribution to the weak mixing phase $\phi_{d,s}$ from selected $B_{d,s} \to PP, VP$ and $VV$ decays.

Some time ago~\cite{apo}, we proposed a test on the value of $\sin \phi_s$ in the
 SM by measuring certain longitudinal branching ratios and CP asymmetries of $B_{d,s}$ mesons decaying into vectors, for modes mediated by penguin diagrams -- we focused on the potential of the $B_s \to \bar{K^{0*}} K^{0*}$ decay at that time.
 The method has advantages both from the theoretical and experimental points of view. 
On the theoretical side, it reduces the required theoretical input mainly to a single hadronic quantity $\Delta$, defined as the difference
between ``tree'' and ``penguin'' contributions. More precisely,
for a $\bar B_Q$ meson decaying through a $b\to q$ penguin-mediated process, the decomposition
\begin{equation}
\bar{A}\equiv A(\bar{B}_Q\to M_1 M_2)
  =\lambda_u^{(q)} T + \lambda_c^{(q)} P\,,
\label{dec}
\end{equation}
with the CKM factors $\lambda_U^{(q)}=V_{Ub} V_{Uq}^*$, can be used to define $\Delta$ as
 the difference of the hadronic matrix elements that are multiplied by $\lambda_u^{(q)}$ and $\lambda_c^{(q)}$ respectively, i.e., 
\eq{\Delta=T-P \label{delta}\ .}
In the case of penguin-mediated decays, the evaluation of this quantity using QCD factorisation (QCDF) is expected to be particularly robust, as it was built specifically to cancel the infrared divergences coming from spectator-quark and annihilation contributions up to next-to-leading order~\footnote{Notice that this cancellation is ensured at NLO in the strong coupling constant, but it has to be proven that this feature remains at higher orders.} \cite{prl}.
On the experimental side, this test of weak mixing angles within the Standard Model can be performed simply by 
measuring a CP-averaged branching ratio and an untagged rate. 
This allows us to avoid tagging, although  a time-dependent analysis is still required.

In the present article, we extend this approach beyond the simple SM test on the weak mixing
angles $\phi_s$ and $\phi_d$ presented in Ref.~\cite{apo}. Indeed, we show that we can also use penguin-mediated decays to measure the size of NP contributions to neutral-meson mixing, assuming that its contribution to $\Delta B=1$ decays is negligible.
We also update our predictions for the branching ratio of $B_s\to \bar{K}^{*0}K^{*0}$ which has been recently measured.
The paper is organized in the following way.  In Section \ref{s1}, we  present a formula to pin down the NP contribution to $\phi_s$ and $\phi_d$,
obtained from $b \to d,s$ transitions, under the assumption that NP provides significant contributions only to the meson-mixing phases. In Section \ref{s2} we discuss the main theoretical input in this method. In Section \ref{s3} we consider three different examples of the method, corresponding to $B \to PP, PV$ and $VV$ decays. We update our prediction in Ref.~\cite{apo} for ${\rm BR^{\rm long}}(B_s \to {\bar K^{*0}} K^{*0})$ in Section \ref{s4}, and compare it with its recent measurement by the LHCb collaboration~\cite{adeva}. In Appendix \ref{A} we provide a dictionary between the experimental measurements~\cite{Aubert:2008zza,belle} and the theoretical quantities~\cite{apo} defined for longitudinal observables in $B\to VV$ decays.  In  Appendix \ref{B}, we discuss differences in the determination of the branching ratio for neutral mesons for tagged or untagged analyses, highlighting the role played by the width difference $\Delta\Gamma$ (a problem particularly relevant for $B_s$ mesons).

\sect{Formulae for NP mixing angles}
\label{s1}

Using the unitarity relation $\lambda_u^{(q)}+\lambda_c^{(q)}+\lambda_t^{(q)}=0$, we can write Eq.~(\ref{dec}) in terms of $\lambda_c^{(q)}$ and $\lambda_t^{(q)}$
\eq{\label{newamp}
\bar{A} \equiv A(\bar{B}_Q\to M_1 M_2)
  =-\lambda_t^{(q)}\, T - \lambda_c^{(q)} \Delta\,.}
 The weak phase in $\lambda_t^{(q)}$ is the angle $\beta_q$, defined as
\eq{\beta_q\equiv \arg \left(- \frac{V_{tb} V_{tq}^*}{V_{cb} V_{cq}^*} \right)= \arg \left(- \frac{\lambda_t^{(q)}}{\lambda_c^{(q)}} \right)\,,}
whereas $\lambda_c^{(q)}$ is real to a very good approximation for both $q=d,s$.  
Following the definitions in Appendix \ref{A}, we introduce the observables: 
a branching ratio $BR = g_{ps}{(|A|^2+|\bar{A}|^2)}/{2}$  (where $g_{ps}$ is the phase space factor)  
and three CP asymmetries $\Adir$, $\Amix$ and $\Adel$
\begin{eqnarray}\label{obs}
\Adir &\equiv&\frac{|A|^2-|\bar{A}|^2}{|A|^2+|\bar{A}|^2},\quad\Amix\equiv -2 \eta_f \frac{{\rm
Im}(e^{-i\phi_Q}A^{*}\bar{A})}{|A|^2+|\bar{A}|^2}\,,\nonumber\\
{A}_{\Delta\Gamma}&\equiv& -2 \eta_f\frac{{\rm Re}(e^{-i\phi_Q}A^{*}\bar{A})}{|A|^2+|\bar{A}|^2}\,,
\end{eqnarray}
defined in terms of $\bar{A}$ and its CP-conjugate $A$, as well as the $B_Q$ meson-mixing phase $\phi_Q$ ($\eta_f$ is the CP parity of the final state). 

These four observables can be written, using Eq.~(\ref{newamp}), in terms of $\lambda_{c,t}^{(q)}$, $T$, $\Delta$ and $\phi_Q$. One can then eliminate the hadronic parameter $T$ to obtain a relationship between these observables
\eqa{2 g_{ps} |\Delta|^2 |\lambda_c^{(q)}|^2\sin^2{\beta_q}&&\label{sr}\\[2mm]
&&\hspace{-2.8cm}=BR(1-\eta_f \sin{\Phi_{Qq}} \Amix+\eta_f \cos{\Phi_{Qq}} \Adel)\,,\nonumber}
%
%
with $\Phi_{Qq}$ defined as
\eq{\Phi_{Qq}=2\beta_Q-2\beta_q+\phi_Q^{\rm NP}\,,\label{phi}}
%
%
being $\phi_Q^{\rm NP}$ the new physics contribution to the mixing angle of the $B_Q$ system: $\phi_Q=2\beta_Q+\phi_Q^{\rm NP}$. In deriving Eq.~(\ref{sr}) we have assumed that New Physics could alter only the mixing phase of the neutral $B_Q$ meson, but not the CKM matrix elements involved in the decay process. 
Eq.~(\ref{sr}) is a generalization of similar formulae developed in Refs.~\cite{prl,apo} in the context of the SM.   This relation  is  the starting point of our analysis.

Collecting all terms on the left-hand side in Eq.~(\ref{sr}), and defining 
\eq{C=\frac{2\,g_{ps}\, |\lambda_c|^2 \sin^2 \beta_q\, |\Delta|^2}{BR},\label{C}}  
we can solve Eq.~(\ref{sr}) for $\Phi_{Qq}$:
\begin{eqnarray} 
\sin\Phi_{Qq}&=&z\ \eta_f{\hat A}_{\rm mix} \pm \sqrt{1-z^2}\ \eta_f{\hat A}_{\Delta \Gamma}\ ,\nonumber\\
\cos\Phi_{Qq}&=&-z\ \eta_f{\hat A}_{\Delta\Gamma} \pm \sqrt{1-z^2}\ \eta_f{\hat A}_{\rm mix}\ .
\label{formulon2}
\end{eqnarray}
Here, $z=(1-C)/\sqrt{1-\Adir^2}$, $\hat{A}_{\rm mix}=A_{\rm mix}/\sqrt{1-\Adir^2}$ and
$\hat{A}_{\Delta \Gamma}=A_{\Delta \Gamma}/\sqrt{1-\Adir^2}$, with the relation:
\eq{\hat{A}_{\rm mix}^2+\hat{A}_{\Delta \Gamma}^2=1\ .}
There is a two-fold ambiguity in Eq.~(\ref{formulon2}). In practice, we will see that $z \simeq 1$ (or equivalently $C\ll1 $), so that the two solutions are very close  and they can be considered as one single solution within current theoretical and experimental uncertainties. Notice also that $z^2 \leq 1$, which provides a very strong and interesting constraint on the size of $\Adir$: 
\eq{
\Adir^2 < C (2 - C)\,.
\label{Adirmax}}
In general, a pair of values for $BR$ and $\Delta$ imply an upper bound on $|\Adir|$~\cite{apo,pro} which is much tighter than the experimental value for $\Adir$. A violation of this bound would be an indication of NP in decay. In the following we will take $\Adir=0\pm \sqrt{2C-C^2}$ in the numerical analyses. For example, in the case of $B \to \phi K_{\rm S}$ we have $C=(8.9\pm 7.7) \times 10^{-5}$, which implies $|\Adir| < 0.019$.

If there is no New Physics and if the process considered is such that $Q=q$, one recovers $\sin \Phi_{Qq} \to 0$ and $\cos \Phi_{Qq} \to 1$, as can be checked
from Eq.~(\ref{phi}). In this case, Eq.~(\ref{sr}) reduces to the simple relation:
\eq{z^{\rm SM}=-\eta_f\hat{A}_{\Delta\Gamma}^{\rm SM}\ .}
This compact equation can be easily rewritten as a SM test of the angles of the unitarity triangle and corresponds to Eqs.~(33)-(36) of Ref.~\cite{apo}.
%
%

\sect{Theoretical Input}
\label{s2}

The theoretical input in this formalism is limited to the quantity $\Delta$ defined in Eq.~(\ref{delta}). As discussed in Refs.~\cite{apo,prl}, this quantity can be computed safely within QCD factorisation for penguin mediated decays because of the cancellation of long-distance contributions. As a consequence of this cancellation, only penguin contractions contribute to $\Delta$,  as can be seen by inspection of the formulae in Ref.~\cite{QCDF}. The general form of $\Delta$ for a $B_Q \to M_1 M_2$ decay is \cite{apo,prl}:
\eq{\Delta_{M_1 M_2}= A^Q_{M_1 M_2} \frac{C_F \alpha_s}{4 \pi N} C_1 \left[ {\bar G}_{M_2}(m_c^2/m_b^2)-{\bar G}_{M_2}(0) \right] \ ,}
where $M_1$ picks up  the spectator quark of the $B_Q$ meson. The normalisation involves the $M_2$ decay constant and a $B\to M_1$ form factor:
\eq{A^Q_{M_1 M_2}= \frac{G_F}{\sqrt{2}} m_{B_Q}^2 f_{M_2} A^{B_Q \to M_1}(0)\ .}
$C_1$ is the relevant Wilson coefficient of the effective Hamiltonian at a scale of order $m_b$. ${\bar G}_{M_2}=G_{M_2} \pm r_{\chi}^{M_2} {\hat G}_{M_2}$ is the penguin function with a $u$ or a $c$ quark running in the loop and the sign depends on whether $M_1$ is a pseudoscalar or a vector meson (see \cite{QCDF,BVV} for the precise definition of the quantities entering $\Delta$). For identical particles in the final state (for instance $B_s \to \phi \phi$) an extra factor of $2$ must be included in $\Delta$.

In the case in which both $M_1$ and $M_2$ are vector mesons, there is a different $\Delta$ for each of the three transversity amplitudes. As discussed in Ref.~\cite{BVV}, 
only longitudinal amplitudes can be handled safely within QCD factorisation, so we shall focus on longitudinal polarisations for vector-vector modes, in the spirit of Ref.~\cite{apo} (see also Ref.~\cite{kagan} for a discussion of transverse amplitudes in QCDF). This requires the definition of suitable longitudinal observables, together with a procedure to extract them from the observables that are actually \emph{measured}, as discussed in more detail in the appendices of the present article.

In Table \ref{table1} we present the values of $\Delta$ for $B_Q \to XY$ penguin-mediated decay channels, with $X,Y=P,V_L$  (where $V_L$ stands for longitudinally polarised vectors), using the same hadronic inputs as in Ref.~\cite{apo}. 

\begin{table}
\begin{tabular}{||l||c|c||}
\hline\hline
\hspace{0.2cm} Channel &$\ |\Delta|\ (10^{-7}\, {\rm GeV})\ $& $\ C\times BR\ $ \\
\hline\hline
$\quad B_d\to K \bar K$  &$(3.23\pm 1.16)$ & $\quad( 29.8\pm 21.9)\cdot 10^{-9}\quad$\\
\hline
$\quad B_s\to \bar K K$  &$(3.05\pm 1.11)$& $\quad( 1.21\pm 0.89)\cdot 10^{-9}\quad$\\
\hline
$\quad B_d\to {K \phi}$  &$(2.32\pm 1.00)$& $\quad( 0.74\pm 0.64)\cdot 10^{-9}\quad$\\
\hline
$\quad B_d\to {K \bar K^*}$ &$(2.29\pm 0.93)$& $\quad( 14.7 \pm 12.1 )\cdot 10^{-9}\quad$\\
\hline
$\quad B_d\to {K^*\bar K}$ &$(0.41\pm 0.60)$& $\quad( 0.47 \pm 1.38 )\cdot 10^{-9}\quad$\\
\hline
$\quad B_s\to {\bar K K^*}$ &$(2.16\pm 0.89)$& $\quad( 0.60 \pm 0.50)\cdot 10^{-9}\quad$\\
\hline
$\quad B_s\to {\bar K^* K}$ &$(0.36\pm 0.53)$& $\quad( 0.02\pm 0.05)\cdot 10^{-9}\quad$\\
\hline
$\quad B_d\to {K^*\bar K^*}$ &$(1.85\pm 0.93)$& $\quad( 9.37\pm 9.53)\cdot 10^{-9}\quad$\\
\hline
$\quad B_s\to {\bar K^*K^*}$  &$(1.62\pm 0.81)$& $\quad( 0.33\pm 0.33)\cdot 10^{-9}\quad$\\
\hline
$\quad B_d\to {K^* \phi}$  &$(1.92\pm 1.03)$& $\quad( 0.49\pm 0.53)\cdot 10^{-9}\quad$\\
\hline
$\quad B_s\to {\phi K^*}$  &$(1.87\pm 0.94)$& $\quad( 8.80\pm 8.96)\cdot 10^{-9}\quad$\\
\hline
$\quad B_s\to {\phi \phi}$  &$(3.86\pm 2.09)$& $\quad( 0.92\pm 1.00)\cdot 10^{-9}\quad$\\
\hline\hline
\end{tabular}
\caption{Values of $\Delta$ for the various decays of interest, and the corresponding values for $C\times BR$, defined in Eq.(\ref{C}). In the case of
two vector mesons these numbers should be understood as referring to longitudinal polarisations (the third column means $C\times BR^{\rm long}$ in this case).}
\label{table1}
\end{table}

\sect{Neutral-meson mixing angles from $B_Q\to PP, PV$ and $V_LV_L$}
\label{s3}

In this section we illustrate the determination of the NP contributions to both weak mixing angles $\phi_d$ and $\phi_s$ using Eqs.~(\ref{sr}) and (\ref{formulon2}). We consider channels of three different types: $B_{s}\to K^0\bar K^0$ ($B\to PP$), $B_d\to \phi K_{\rm S}$ ($B\to PV$), and $B_d\to \phi K^*$ ($B\to V_LV_L$). We will see that in the case of $B_d\to \phi K_{\rm S}$ and $B_d\to \phi K^*$, all the required observables are already measured, which allows us to perform a complete analysis to extract $\phi_d^{\rm NP}$. There is no experimental data for $B_{s}\to K^0\bar K^0$, so we will only be able to present a case study, which can be exploited immediately as soon as data is available.

Besides the theory input $\Delta$ and the measured branching ratio, we need additional inputs for the SM CKM parameters. Since we are assuming that meson mixing could be affected by NP, we must use values of the CKM elements extracted using modes insensitive to mixing.  The fit in Table 11 of Ref.~\cite{CKM} satisfies this requirement, as the mixing-related observables included in this fit are used essentially to determine the size of NP in meson mixing and have a very small impact on the CKM parameter themselves, which are thus determined from tree-dominated quantities. Indeed we checked that a fit limited to the tree dominated inputs of Ref.~\cite{CKM} yields almost identical results for the CKM parameters. From Ref.~\cite{CKM} we obtain, upon symmetrizing errors:
\eqa{
|\lambda_u^{(d)}| & = & (4.1\pm 0.3)\cdot 10^{-3}\,,\nonumber\\
|\lambda_u^{(s)}| &=&  (9.5\pm 0.6)\cdot 10^{-4}\,, \nonumber\\
|\lambda_c^{(d)}| & = & (9.2\pm 0.2)\cdot 10^{-3}\,,\nonumber\\
|\lambda_c^{(s)}|  &=&  0.041\pm 0.001\,,\label{SMfit}\\
\beta_d & = & (26.2\pm 2.1)^\circ\,,\nonumber\\
\beta_s &=&  -(1.26\pm 0.07)^\circ\,,\nonumber\\
\gamma & = & (69.9\pm 4.4)^\circ\,,\nonumber
}
which will be used in the following studies.

\subsection{First example: $B_d\to \phi K_{\rm S}$}

\begin{figure}
\includegraphics[width=8cm]{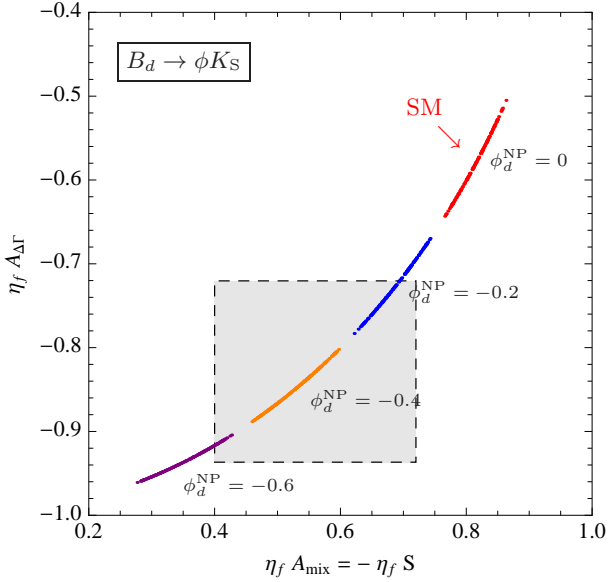}
\Text(-185,190)[lb]{\framebox{$B_d\to\phi K_{\rm S}$}}
\Text(-160,33)[lb]{\scriptsize $\phi_d^{\sss \rm NP}=-0.6$}
\Text(-112,65)[lb]{\scriptsize $\phi_d^{\sss \rm NP}=-0.4$}
\Text(-75,106)[lb]{\scriptsize $\phi_d^{\sss \rm NP}=-0.2$}
\Text(-44,156)[lb]{\scriptsize $\phi_d^{\sss \rm NP}=0$}
\Text(-83,165)[lb]{\color{red} \begin{minipage}{1cm} SM\\ $\qquad\searrow$\end{minipage}}
\caption{$\Amix$ vs. $\Adel$, for several values of the NP mixing angle $\phi_d^{\sss \rm NP}$ for $B_d\to \phi K_{\rm S}$, where $BR=(8.3\pm 1.1)\cdot 10^{-6}$, $\Delta=(2.32\pm1.00)\cdot 10^{-7} {\,\rm GeV}$ and $g_{ps}=8.4\cdot 10^9{\,\rm GeV^{-2}}$. The box indicates the experimental value for the asymmetry $\Amix=-0.56\pm0.16$ and the range obtained for $A_{\Delta \Gamma}$ using $\Adir=0\pm0.019$ ($\eta_f=-1$).}
\label{fig1}
\end{figure}

We first consider $B_d\to\phi K_{\rm S}$. Using the formulae in Eq.~(\ref{formulon2}), together with $\Delta$ and $C$ in Table~\ref{table1}, and the experimental numbers $BR=(8.3\pm 1.0)\cdot 10^{-6}$ and $\Amix=-0.56\pm0.16$~\cite{HFAG} (the product of intrinsic parities is $\eta_{\phi K_{\rm S}}=-1$), we find the two solutions for the NP contribution to the mixing angle:
\eq{\phi_d^{\rm NP}=(-0.38\pm 0.21) \vee (-0.35\pm 0.21)\ {\rm rad}\,,}
from which we can give an averaged result:
\eq{\phi_{d,\,{\rm aver}}^{\rm NP}=-0.36\pm 0.22\ {\rm rad}\ .
\label{res1}}
This result is (marginally) compatible with the SM value $\phi_d^{\rm NP}=0$. However the error on $\phi_d^{\rm NP}$ is almost completely dominated by the experimental uncertainty in $\Amix$: any improvement on the latter would impact our knowledge of $\phi_d^{\rm NP}$ and its agreement with SM expectations.

In Figure \ref{fig1} we show the regions in the $\Amix$-$\Adel$ plane corresponding to different values of $\phi_d^{\rm NP}$, the NP contribution to the mixing angle. The easiest way to understand this plot is through the two constraints on $\Amix$ and $\Adel$: Eq.~(\ref{sr}) is a linear equation at fixed $\phi_Q^{\rm NP}$, whereas
%
%
%
$\Amix^2+\Adel^2\le1$ yields a radial constraint. For a fixed $\phi_Q^{\rm NP}$, the solution for $\Amix$ and $\Adel$ is thus given by the intersection of a line with the unit circle. If this line is tangent, the solution is limited to \emph{one} point, and our determination is perfectly precise. The opposite situation occurs if the line goes through the origin. It turns out that the distance of this line from the origin is given exactly by $1-C$,  so $C$ measures the precision with which we can determine $\phi_Q^{\rm NP}$:\\[-2mm]

\noindent \emph{The smaller the value of $C$, the more precisely we can pin down the value of $\phi_Q^{NP}$}.\\[-2mm]

The SM solution is shown explicitly in Fig.~\ref{fig1}, and the gray box indicates the current experimental situation. One can see there is marginal agreement, in line with the result in Eq.~(\ref{res1}). We stress that these results include all known hadronic uncertainties, and the errors are relatively small because the chosen theoretical input is robust. A more precise value for the angle $\beta_d$ would achieve a substantial reduction of the regions of fixed $\phi_d^{\rm NP}$ in Fig.~\ref{fig1}.

It should be mentioned that, since we have no way of determining the sign of $A_{\Delta\Gamma}$, there is a second box in Fig.~\ref{fig1} that has not been drawn, corresponding to $\eta_f A_{\Delta\Gamma}\sim +0.8$. We have discarded this possibility because it leads to a very large New Physics mixing angle.

\subsection{Second example: $B_d\to \phi K^*$}

\begin{figure}
\includegraphics[width=8cm]{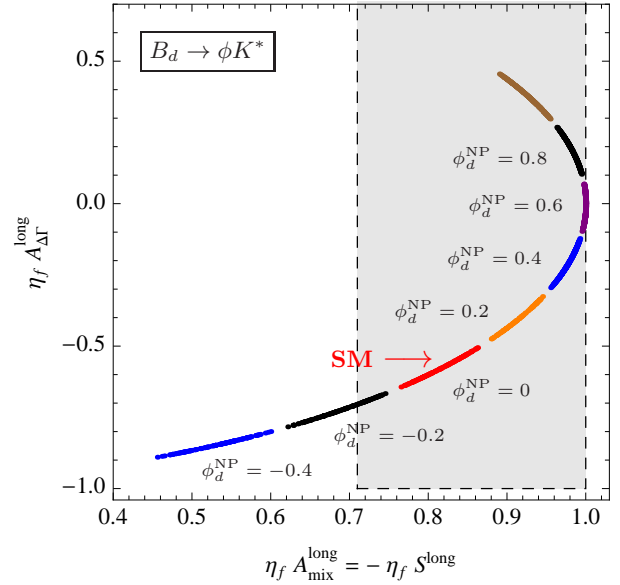}
\Text(-178,195)[lb]{\framebox{$B_d\to\phi K^*$}}
\Text(-105,53)[lb]{\scriptsize $\phi_d^{\sss \rm NP}=-0.2$}
\Text(-60,70)[lb]{\scriptsize $\phi_d^{\sss \rm NP}=0$}
\Text(-82,100)[lb]{\scriptsize $\phi_d^{\sss \rm NP}=0.2$}
\Text(-62,120)[lb]{\scriptsize $\phi_d^{\sss \rm NP}=0.4$}
\Text(-54,140)[lb]{\scriptsize $\phi_d^{\sss \rm NP}=0.6$}
\Text(-60,158)[lb]{\scriptsize $\phi_d^{\sss \rm NP}=0.8$}
\Text(-155,40)[lb]{\scriptsize $\phi_d^{\sss \rm NP}=-0.4$}
\Text(-106,83)[lb]{\bf \color{red} SM \large $\longrightarrow$}
\caption{$\Amix$ vs. $\Adel$, for several values of the NP mixing angle $\phi_d^{\sss \rm NP}$for  $B_d\to \phi K^*$, where $BR^{\rm long}=(4.7\pm 0.4)\cdot 10^{-6}$, $\Delta=(1.92\pm1.03)\cdot 10^{-7}{\,\rm GeV}$ and $g_{ps}=8.2\cdot 10^9{\,\rm GeV^{-2}}$. The box indicates the experimental values for the asymmetries given in Eq.~(\ref{obslong}) ($\eta_f=+1$).}
\label{fig2}
\end{figure}

An angular analysis of the decay $B_d\to \phi K^*$ is available from both Babar and Belle collaborations~\cite{Aubert:2008zza,belle}, with an additional time-dependent analysis for the former experiment. The averaged results read \cite{HFAG,Aubert:2008zza}
\eq{
\hspace{-0.05cm}
\begin{array}{rclrcl}
BR & = & (9.8 \pm 0.7) \cdot 10^{-6}\,, &\quad {\cal A}_{CP} &=&  0.01\pm 0.05\,,\\
f_L & = & 0.48\pm 0.03\,, & {\cal A}^0_{CP} &=&  0.04\pm 0.06\,,\\
\Delta\phi_0 & = & 0.28 \pm 0.42\,, & \Delta\delta_0 &=&  0.27 \pm 0.16\,.
\end{array} \hspace{-0.12cm}
}
Using equations (\ref{brL})-(\ref{AdL}) in Appendix \ref{A} (and noting that $\eta_{\phi K^*}=+1$), we obtain the longitudinal observables:
\eqa{
BR^{\rm long} & = & (4.7 \pm 0.4) \cdot 10^{-6}\,,\nonumber\\
A_{\rm dir}^{\rm long} & = & -0.05\pm 0.08\,,\nonumber\\
A_{\rm mix}^{\rm long} & = & 0.96\pm 0.25\,,\label{obslong}\\
A_{\Delta\Gamma}^{\rm long} & = &\pm( 0.27\pm 0.86)\,.\nonumber
}
Compared to the case of $B_d\to \phi K_{\rm S}$, the uncertainty on $A_{\Delta\Gamma}^{\rm long}$ is so large that one cannot distinguish, among the two solutions, which one is disfavored by too
large a value of the NP mixing angle. For illustration we focus on the negative solution of $\Adel^{\rm long}$.

For this decay, Eq.~(\ref{Adirmax}) yields the following range for the direct asymmetry:
\eq{A_{\rm dir}^{\rm long}=0 \pm 0.015\ ,}
which is more precise than the experimental value in Eq.~(\ref{obslong}), and which is used in the following. Equation (\ref{formulon2}) then yields the two solutions:
\eqa{\phi_d^{\rm NP}&=&(0.31\pm 0.90) \vee (0.34\pm 0.90)\ {\rm rad}\ ,}
which can be averaged as
\eq{\phi_{d,\,{\rm aver}}^{\rm NP}=0.33\pm 0.90\ {\rm rad}\,.
\label{res2}}
This result is compatible with the SM ($\phi_d^{\rm NP}=0$), and also with the result obtained from $B_d\to \phi K_{\rm S}$ (Eq.~(\ref{res1})), within large uncertainties.

In Fig.~\ref{fig2}, we show the regions in the $\Amix$-$\Adel$ plane corresponding to different values of the NP contribution to the mixing angle. The SM solution is shown explicitly, and the gray box indicates the current experimental situation. From this plot we see that a more accurate knowledge of $A_{\rm mix}^{\rm long}$ or $\Adel^{\rm long}$ would be very useful to test the SM hypothesis.

\subsection{Third example: $B_{s}\to K^0 \bar K^0$}

There is currently no information on this mode. However, the U-spin related decay $B_d\to K^0 \bar K^0$ has the measured branching ratio~\cite{HFAG}:
\eq{BR(B_d\to K^0 \bar K^0)=(0.96 \pm 0.20)\cdot 10^{-6}}
As explained in Ref.~\cite{prl}, if there is no NP affecting the decay $B_d\to K^0 \bar K^0$, one can predict the SM branching ratio for the decay $B_s\to K^0 \bar K^0$ quite precisely. This prediction is not affected by a possible NP contribution to $\phi_d$ (see Section 3 of Ref.~\cite{apo}), and since we are assuming that there is no NP affecting $\Delta B=1$ processes, it is a good option to take this prediction as an input for our analysis~\cite{prl}:
\eq{BR(B_s\to K^0 \bar K^0)=(18.2 \pm 7.3)\cdot 10^{-6}\,.}
The present analysis of this decay should be considered as a case study, waiting for data to be available.

\begin{figure}
\includegraphics[width=8cm]{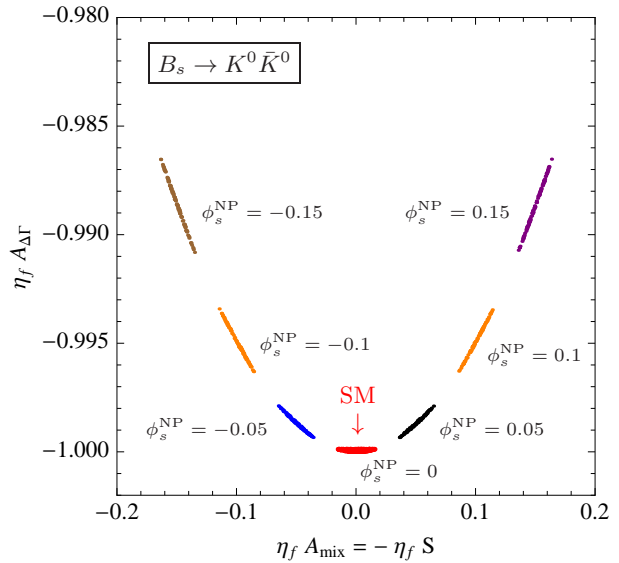}
\Text(-175,183)[lb]{\framebox{$B_s\to K^0 \bar K^0$}}
\Text(-95,32)[lb]{\scriptsize $\phi_s^{\sss \rm NP}=0$}
\Text(-65,48)[lb]{\scriptsize $\phi_s^{\sss \rm NP}=0.05$}
\Text(-47,76)[lb]{\scriptsize $\phi_s^{\sss \rm NP}=0.1$}
\Text(-78,130)[lb]{\scriptsize $\phi_s^{\sss \rm NP}=0.15$}
\Text(-176,48)[lb]{\scriptsize $\phi_s^{\sss \rm NP}=-0.05$}
\Text(-133,80)[lb]{\scriptsize $\phi_s^{\sss \rm NP}=-0.1$}
\Text(-155,130)[lb]{\scriptsize $\phi_s^{\sss \rm NP}=-0.15$}
\Text(-110,50)[lb]{\color{red} \begin{minipage}{1cm} SM\\ $\downarrow$\end{minipage}}
\caption{$\Amix$ vs. $\Adel$, for several values of the NP mixing angle $\phi_s^{\sss \rm NP}$ for $B_s\to K^0 \bar K^0$, where we take the estimate $BR=(18.2\pm 7.3)\cdot 10^{-6}$ (see the text). In addition, we used $\Delta=(3.05\pm1.11)\cdot 10^{-7}{\,\rm GeV}$  and $g_{ps}=8.03\cdot 10^9{\,\rm GeV^{-2}}$.}
\label{fig3}
\end{figure}

Without further experimental information on CP asymmetries, we cannot use Eq.~(\ref{formulon2}) to extract $\phi_s^{\rm NP}$, the NP contribution to the mixing angle. But it is still possible to determine the regions in the $\Amix$-$\Adel$ plane that correspond to each value of $\phi_s^{\rm NP}$, similarly to Figs.~\ref{fig1} and \ref{fig2}. 
Notice that in the present case we have simply $\Phi_{Qq}=\phi_s^{\rm NP}$, since $Q=q=s$ here. The case $B_s\to K^0 \bar K^0$ is shown in Figure \ref{fig3}. Whenever the branching ratio for this mode is measured, this plot can be remade, although we do not expect it to change in an appreciable way. A precise measurement of the time-dependent CP asymmetry will then provide an accurate determination of $\phi_s^{\rm NP}$.

One may wonder why we did not consider the decay $B_d\to K^0 \bar K^0$ at a first stage, since its branching ratio is known. However, the prospects for this particular decay are not very alluring. The mixing angle involved in the expression for $C$ (Eq.~(\ref{C})) is $\beta_d$, and the branching ratio is not particularly large, so that the value of $C$ is small but not tiny ($C\sim 0.03$). As mentioned above, the value of $C$ determines the potential accuracy of our procedure, which turns out to be rather poor in the present case.  This is true in general for all $b\to d$ transitions in our list (see Table~\ref{table1}); however, the situation may change if the branching ratio of some of the modes is found to be large enough, as it might be for $B_d\to K^{*0} \bar K^{*0}$.

The three plots in Figs.~\ref{fig1},\ref{fig2} and \ref{fig3} can be reversed and reinterpreted as predictions for the $\Amix$ and $A_{\Delta \Gamma}$ asymmetries if the range for $\phi_{d,s}^ {NP}$ is extracted from other processes, always under the assumption that there is no significant New Physics contribution to $\Delta B=1$ processes.

\sect{The Branching ratios $BR(B_{d,s} \to {\bar K}^{*0} K^{*0})$ within the SM}
\label{s4}

In this section, we present a prediction for $BR^{\rm long}(B_s \to {\bar K}^{*0} K^{*0})$ in the SM using our approach.
This prediction can be easily turned, as shown later on, into a prediction for the total branching ratio, once $f_L$, ${\cal A}_{CP}$ and ${\cal A}_{CP}^0$ are measured (see Appendix \ref{A} for details). This section is an update of results presented in Ref. \cite{apo}.

The two decays $B_q \to {\bar K}^{*0} K^{*0}$, with $q=d,s$, are related by U-spin symmetry. The symmetry-breaking effects can be separated into a factorizable and a non-factorizable part. This translates into the following relations involving the hadronic parameters $T_q$ and $P_q$
\begin{equation}
P_s  =  f\,P_d(1+\delta_P)\ ,\quad  T_s  =  f\,T_d(1+\delta_T)\,,
\end{equation}
with the factorisable factor $f$
\begin{equation}
f=\frac{m_{B_s}^2A_0^{B_s\to K^*}}{m_B^2A_0^{B\to K^*}}=\frac{\Delta^s_{K^*K^*}}{\Delta^d_{K^*K^*}}\ ,
\end{equation}
and $\delta_{T,P}$ account for the non-factorizable symmetry breaking effects. There is no theoretically clean way to compute these quantities. Here we use QCD-factorisation to \emph{estimate} an upper bound on this corrections. We get
$|\delta_P|<0.09$, and keep its phase as a free parameter (even though QCD factorisation would predict it to be small). It is very easy to show that the quantity $\delta_T$ is related to $\delta_P$ through
\begin{equation}\label{TdPd}
T_d\, \delta_T=P_d\, \delta_P\,.
\end{equation}

Observables related to the decay $B_s \to {\bar K}^{*0} K^{*0}$ can be obtained if the ones for  $B_d \to {\bar K}^{*0} K^{*0}$ are known. An important result discussed in Ref.~\cite{apo} is that the ratio
\begin{equation}
R_{sd}\equiv\frac{BR^{\rm long}(B_s\to K^{*0}\bar{K}^{*0})}{BR^{\rm long}(B_d\to K^{*0}\bar{K}^{*0})}\,,
\label{defRsd}
\end{equation}
is almost a constant if $BR^{\rm long}(B_d\to K^{*0}\bar{K}^{*0})\gtrsim 10^{-7}$, even in the presence of NP in mixing. Indeed, when  the two branching ratios in Eq.~(\ref{defRsd}) are reexpressed in terms of $P_{s,d}$ and $\Delta_{s,d}$ (as computed in Table~\ref{table1}), it becomes clear that the ratio $R_{sd}$ is completely dominated by the penguin contributions $P_{s}$ and $P_{d}$ for large enough $BR^{long}(B_d\to K^{*0}\bar{K}^{*0})$, and thus determined by $f$ and $\delta_P$ essentially.
This behaviour is confirmed in Table \ref{tabRsd} for several values of the branching ratio.

\begin{table}
\begin{tabular}{||c||c|c|c||}
\hline\hline
$BR_d$ & $5\cdot 10^{-7}$ & $5\cdot 10^{-6}$ & $5\cdot 10^{-5}$ \\
\hline\hline
$R_{sd}^{\rm \sss DMV}$ & $16.05\pm4.87$ & $16.38\pm4.92$ & $16.46\pm4.93$\\[1mm]
\hline\hline
\end{tabular}
\caption{Results for the ratio $R_{sd}^{\rm \sss DMV}$ for three different values of $BR^{long}(B_d\to K^{*0}\bar{K}^{*0})$. It can be seen that the dependence on this branching ratio is very mild.}
\label{tabRsd}
\end{table}

Our result, combining all the error sources, is
\begin{equation}\label{RsdDMV}
R_{sd}^{\rm \sss DMV}=16.4\pm5.2\ ,
\end{equation}
which updates Table IV in Ref.~\cite{apo}. The improvement on the uncertainty comes in particular from Eq.~(\ref{TdPd}) which was not used in this reference.  We have added the superscript ``DMV'' to distinguish our determination from other ones that we will describe now.

Indeed this number can be compared with the value obtained within the QCDF framework taking the usual model for $1/m_b$-suppressed corrections described in Ref.~\cite{BVV}. The corresponding predictions are for the $B_s$ decay mode: $BR(B_s \to {\bar K}^{0*} K^{0*})=(9.1^{+11.3}_{-6.8}) \times 10^{-6}$ and $f_L(B_s)=0.63^{+0.42}_{-0.29}$,
and for the corresponding $B_d$ decay mode: $BR(B_d \to {\bar K}^{0*} K^{0*})=(0.6^{+0.5}_{-0.3}) \times 10^{-6}$ and $f_L(B_d)=0.69^{+0.34}_{-0.27}$. 
We write $R_{sd}$ in terms of total branching ratios and polarisation fractions, as described in Appendix \ref{A}
\eq{R_{sd}= \frac{BR(B_s) f_L(B_s)}{BR(B_d) f_L(B_d)}\, f_c\,, }
where we introduce the correcting factor
\eq{f_c= \frac{ 1 + {\mathcal A}_{CP}^0 (B_s)  {\mathcal A}_{CP} (B_s) }{ 1 +  {\mathcal A}_{CP}^0 (B_d)  {\mathcal A}_{CP} (B_d)}\ .}
In Ref.~\cite{BVV}, the predictions within QCDF are $ {\mathcal A}_{CP}(B_s) \simeq 1\%$ and $ {\mathcal A}_{CP}(B_d) \simeq -13\%$. Since the QCDF predictions are dominated by longitudinal polarisation, we assume $ {\mathcal A}_{CP}^0(B_s)$ and  $ {\mathcal A}_{CP}^0(B_d)$ to be of the same size, leading to a factor $f_c$ ranging from 0.98 to 1, and to the QCDF prediction
\eq{R_{sd}^{\rm QCDF-I}=13.8 \pm 19.2\ .
\label{RsdQCDFI}}                   
Alternatively, if data on $B \to K^* \phi$ is used to control annihilation rather than the usual model for these $1/m_b$-suppressed corrections, the predictions become \cite{BVV}: 
$BR(B_s \to {\bar K}^{0*} K^{0*})=(7.9^{+4.3}_{-3.9}) \times 10^{-6}$, $f_L(B_s)=0.72^{+0.16}_{-0.21}$,
$BR(B_d \to {\bar K}^{0*} K^{0*})=(0.6^{+0.3}_{-0.2}) \times 10^{-6}$ and $f_L(B_d)=0.69^{+0.16}_{-0.20}$. In this case the prediction is practically the same  but with smaller errors:
\eq{R_{sd}^{\rm QCDF-II}=13.7 \pm 10.5\,,}
which is consistent with our result because of the large error bars, even though the central value is a bit low compared to ours.

The most recent experimental values~\cite{HFAG} for $BR(B_d \to {\bar K}^{0*} K^{0*})=(1.28^{+0.35}_{-0.30} \pm 0.11) \times 10^{-6}$ and $f_L(B_d)=0.80^{+0.10}_{-0.12} \pm 0.06\ $ are consistent with the QCDF prediction, although both are  on the high range. 
The LHCb collaboration has measured the $B_s$ mode very recently, reporting the following numbers \cite{adeva}: $BR(B_s \to {\bar K}^{0*} K^{0*})=(2.81 \pm 0.46 \pm 0.45 \pm 0.34) \times 10^{-5}$ and $f_L(B_s)=0.31 \pm 0.12 \pm 0.04$. 

Following the discussion in Appendix \ref{B}, we can express the ratio $R_{sd}$ in terms of experimentally measured quantities:
\eq{R_{sd}=\frac{BR_{LHCb}(B_s) f_{L,LHCb}(B_s)}
{BR_{B-fact}(B_d) f_{L,B-fact}(B_d)}\frac{1-y^2}{1+y\cos\phi_s} ,}
where we define $y=\Delta\Gamma_s/(2\Gamma_s)$. In both cases, it was assumed in the experimental analysis that there is no $CP$-violation in decay, and therefore we do not include the corrections associated with the direct asymmetries. We take $y=0.046\pm 0.027$ \cite{HFAG}, and the SM value of $\phi_s$. These results imply a quite low value for $R_{sd}$:
\eq{R_{sd}^{exp-I}=(8.1\pm 3.3 )\times \left(\frac{f_L(B_s)}{0.31} \right)= 8.1 \pm 4.7\ .
\label{RsdexpI}}
The measured value of $f_L(B_s)$ is unexpectedly low with respect to the polarisation fraction for the $U$-spin related channel: $f_L(B_d) \simeq 0.8$. If $f_L(B_s) \simeq f_L(B_d)$, we would obtain a ratio $R_{sd} \simeq 21$ in better agreement with Eq.~(\ref{RsdDMV}), although $f_L(B_s)\simeq 0.6$ would fit better. 

An alternative determination of this branching ratio using data from $B_s^0 \to D_s^{\pm} \pi^{\mp}$ yields the result \cite{LHCbNote}
 $BR(B_s \to {\bar K}^{0*} K^{0*})=(2.64 \pm 0.61 \pm 0.42) \times 10^{-5}$ in fair agreement with the previous result:
\eq{R_{sd}^{exp-II}=(7.6 \pm 3.2) \times \left(\frac{f_L(B_s)}{0.31} \right)= 7.6 \pm 4.5\ .}
A similar exercise with $f_L(B_s) \simeq 0.8$ yields $R_{sd} \simeq 19$, once again in better agreement with our expectations. 
 
 Finally, one can invert these relations and get a prediction for the total $BR(B_s \to {\bar K}^{0*} K^{0*})$, as measured at a hadronic machine (thereby the subscript `LHCb'):
 \eqa{
&&BR(B_s \to {\bar K}^{0*} K^{0*})_{LHCb}^{\rm \sss DMV}=\nonumber\\[2mm]
&&\hspace{-0.3cm}=R_{sd}^{\rm \sss DMV} \, BR_{B-fact}(B_d) \, \frac{f_{L,B-fact}(B_d)}{f_{L,LHCb}(B_s)} \frac{1+y\cos{\phi_s}}{1-y^2}\nonumber\\[2mm]
&&\hspace{-0.3cm}=(5.7\pm 2.5 )\cdot 10^{-5} \times \left(\frac{0.31}{f_L(B_s)} \right)= (5.7 \pm 3.4)\cdot 10^{-5}\ ,\nonumber\\
 } 
 which can be trivially updated if the longitudinal polarisation of the $B_s$ mode changes.
 
 We remind the reader that, as opposed to the rest of the paper, all the predictions in this section have been obtained within the SM. A detailed analysis of the impact of NP on these observables is certainly worthwhile, but lies beyond the scope of the present article.
  
\section*{CONCLUSIONS}

We have shown how to exploit $B \to PP, PV$ and $VV$ penguin-mediated decays in order to obtain the New Physics contribution to the weak mixing phase of both $B_d$ and $B_s$ systems, under the assumption of no significant New Physics contribution to the decay amplitudes. The main theoretical input consists of the infrared-safe quantity $\Delta$, which can be evaluated within QCD factorisation with a good theoretical control. It is interesting that the experimental knowledge of the branching ratio by itself yields a stringent bound on the direct CP asymmetry, Eq.~(\ref{Adirmax}). As a consequence, the knowledge of BR and $\Amix$ is sufficient to determine the New Physics contribution to the weak mixing angles. Different examples are discussed and the corresponding values for $\phi_Q^{\rm NP}$ presented;  however,  more precise data is required before a clear conclusion can be reached.

\begin{figure}
\psfrag{A}{\hspace{-0.5cm}$\phi_d^{\rm NP}$}
\psfrag{DDDDDDDDD}{\hspace{0.8cm}$(c\bar c)$}
\psfrag{EEEEEEEEE}{\hspace{-0.5cm} Scen. I \cite{CKM}}
\psfrag{CCCCCCCCC}{\hspace{-0.2cm}$B_d\to \phi K$}
\psfrag{BBBBBBBBB}{\hspace{-0.2cm}$B_d\to \phi K^*$}
\includegraphics[width=8cm]{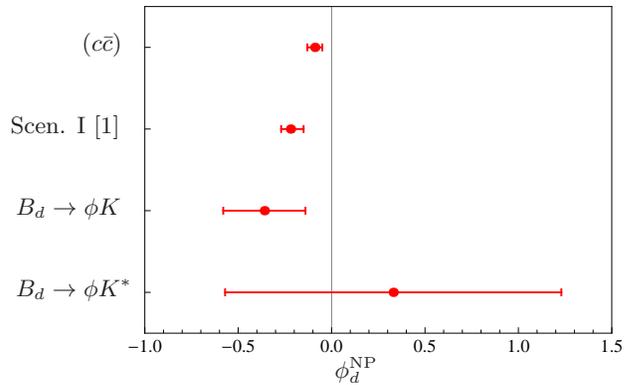}
\caption{Comparison between the NP mixing angles obtained from $c\bar c$, from $B_d\to\phi K_{\rm S}$, from $B_d\to \phi K^*$, and from the fit in Scenario I of Ref.~\cite{CKM}. The line $\phi_d^{\rm NP}=0$ corresponds to the Standard Model.}
\label{fig4}
\end{figure}

These results should be compared with the one obtained from charmonium modes, where one is also sensitive to the NP mixing angle.  Time-dependent analyses of $B\to J/\Psi K_{\rm S}$ give the average \cite{CKM,HFAG}: $\phi_d^{c \bar c}=(21.2\pm 0.9)^\circ$. Using the value quoted in Eq.~(\ref{SMfit}) for the SM contribution $\beta_d$ leads to the estimate
\eq{\phi_d^{\rm NP}(c\bar c)=-0.09\pm 0.04\ {\rm rad}\ .}
In Ref.~\cite{CKM}, the analysis of NP contributions to both $B_d$ and $B_s$ mixings (so-called scenario I) including the experimental information available at that time led to the following value for the NP mixing phase in the $B_d$ sector
\eq{\phi_d^{\rm NP}([1])=-0.22^{+0.07}_{-0.05}\ {\rm rad}\ .}
The main ingredient for this value is the discrepancy between $\phi_d^{c\bar{c}}$ and $BR(B\to\tau\nu)$ in the SM, and thus it will only be marginally affected by the CDF, D{\O} and LHCb updates concerning the $B_s$ sector.
The comparison of these numbers with the ones obtained in Section \ref{s3} is shown in Fig.~\ref{fig4}. 

Finally, an updated result for the longitudinal observables of the decay mode $B_s \to {\bar K^{0*}} K^{0*}$  has been presented. For the ratio of $B_s$ to $B_d$ longitudinal branching ratios, defined in Eq.~(\ref{defRsd}), we find:
\begin{equation}
R_{sd}^{\rm \sss DMV}=16.4\pm5.2\,.
\end{equation}
This result can be compared with recent experimental analyses, and with similar theoretical predictions obtained with alternative approaches:
\begin{enumerate}
\item Our result is compatible with the QCDF prediction (see Eq.~(\ref{RsdQCDFI})), although our error is about four times smaller. If the QCDF is supplemented with $B_d\to K^*\phi$ data to fix $1/m_b$-suppressed corrections of phenomenological relevance, the error decreases, but still is twice as large as ours.
\item Our result is compatible with the recent LHCb measurement,  Eq~(\ref{RsdexpI}), although the experimental number seems rather low. We tentatively identify this anomaly with the surprisingly low value for the longitudinal polarisation fraction $f_L(B_s)\simeq 0.3$: a value of $f_L\simeq 0.8$, similar to the one measured for the $U$-spin related $B_d$ mode, would lead to a $R_{sd}\simeq 20$, more in agreement with our prediction. If the LHCb measurement of $f_L(B_s)$ is confirmed, its interpretation would constitute a theoretical challenge, as it would require a type of NP that violates the flavour symmetry relating $B_d$ and $B_s$ decays  very significantly.
\end{enumerate}

We hope that these results will trigger more precise experimental analyses of penguin-mediated decays both in $B_d$ and $B_s$ sectors, considering the potential of these channels to identify New Physics in meson mixing.

\bigskip

\begin{acknowledgments}
We would like to thank B.~Adeva for discussions and comments. J.V. is supported in part by ICREA-Academia funds. J.M. acknowledges financial support from FPA2008-01430, SGR2009-00894.
\end{acknowledgments}

\appendix

\section{Dictionary for longitudinal observables in $B\to VV$ decays}
\label{A}

In the case of two-body $B$ decays into vector mesons $V_1$ and $V_2$, there are three different configurations of helicity available $h=+1,0,-1$. This is described by three transversity amplitudes $A_{0,\perp,||}$ corresponding to linearly polarised states, with definite CP properties ($A_{0,||}$ correspond to a $CP$-parity $\eta_{0,||}=\eta_{V_1}\eta_{V_2}$, whereas $A_{\perp}$ has $\eta_{\perp}=-\eta_{V_1}\eta_{V_2}$). As discussed extensively in Ref.~\cite{kagan,BVV}, the transversal amplitudes $A_{\perp,||}$ are suppressed by powers of $1/m_B$ compared to the longitudinal amplitude $A_0$, leading to an expected longitudinal polarisation $f_L$ close to 1. In turn, this implies that predictions based on the heavy-quark limit, such as those from QCD factorisation or SCET, are much more reliable for longitunal observables than for parallel and perpendicular ones, where only rough estimates can be derived.

This theoretical situation explains why we considered only longitudinal observables for $VV$ modes in Ref.~\cite{apo} as well as in the present article. 
We defined longitudinal observables as the CP observables constructed by considering as final CP eigenstate the two-meson state with longitudinal polarisations, i.e., the CP-averaged branching ratio and the CP-asymmetries 
\begin{eqnarray}\label{eq:br}
BR^{\rm long}&=&g_{ps}\frac{|A_0|^2+|\bar{A}_0|^2}{2}\,,\\\label{eq:adir}
\Adir^{\rm long} &=& \frac{|A_0|^2-|\bar{A}_0|^2}{|A_0|^2+|\bar{A}_0|^2}\,,\\\label{eq:amix}
A_{\Delta\Gamma}^{\rm long}+ i \Amix^{\rm long} 
   &=& -2\eta_0 \frac{e^{-i\phi_{Q}}A_0^*\bar{A}_0}{|A_0|^2+|\bar{A}_0|^2}\,,
\end{eqnarray}
with the phase space factor relating an amplitude to the corresponding branching ratio
\eq{
g_{ps}=\frac{\tau_{B}\sqrt{[m_{B}^2-(m_{V_1}+m_{V_2})^2][m_{B}^2-(m_{V_1}-m_{V_2})^2]}}{16\pi m_B^3}\,.
}
$A_0$ represents the longitudinal decay amplitude for $B_Q$, $\bar{A}_0$ its CP conjugate, and $\phi_Q$ the meson-mixing angle.

In principle, these quantities can be obtained from the angular analysis of the differential decay width (with respect to the usual angular variables $\theta_1,\theta_2,\psi$ \cite{BVV})
\begin{eqnarray}\label{angularwidth}
&&\hspace{-1.5cm}\frac{d^3\Gamma_{B_q \to V_1V_2}}{d\cos\theta_1d\cos\theta_2d\psi}=
\frac{9}{8\pi}[|a_0|^2\cos^2\theta_1\cos^2\theta_2\nonumber\\
&&+|a_{||}|^2\frac{1}{2}\sin^2\theta_1\sin^2\theta_2\cos^2\psi\nonumber\\
&&+|a_\perp|^2\frac{1}{2}\sin^2\theta_1\sin^2\theta_2\sin^2\psi\nonumber\\
&&+{\rm Re}[a_0^*a_{||}]\frac{1}{2\sqrt{2}}\sin 2\theta_1 \sin2\theta_2 \cos\psi\nonumber\\
&&-{\rm Im}[a_0^*a_{\perp}]\frac{1}{2\sqrt{2}}\sin 2\theta_1 \sin2\theta_2 \sin\psi\nonumber\\
&&-{\rm Im}[a_{||}^*a_{\perp}]\frac{1}{2}\sin^2\theta_1 \sin^2\theta_2 \sin2\psi]\,.
\end{eqnarray}
If the initial meson is charged, or if we neglect the difference of lifetimes between neutral mesons ($\Delta \Gamma=0$), the coefficients $a_i$ and the transversity amplitudes $A_i$ can be identified up to a global normalisation factor, and the quantities (\ref{eq:br})-(\ref{eq:amix}) are easy to obtain from an angular analysis of the decay, as explained in the following. 

For the moment, the only available detailed time-dependent analyses for $VV$ modes of interest to us concerns $B_d\to \phi K^*$~\cite{Aubert:2008zza,belle}, providing CP-averaged parameters and CP-asymmetries for the 3 transversity amplitudes (these analyses for the $B_d$ meson set $\Delta\Gamma_{B_d}=0$). 
Using a $-$ superscript for $B_d$ observables and a $+$ superscript for $\bar{B}_d$ observables, one has values for the branching ratios
\begin{eqnarray}
BR^+&=&\frac{ \bar \Gamma}{\Gamma_{\rm total}}=g_{ps} \sum_\lambda |\bar A_\lambda|^2, \cr BR^-
&=&\frac{\Gamma}{\Gamma_{\rm total}}=g_{ps} \sum_\lambda |A_\lambda|^2\,,
\end{eqnarray}
and for the polarisation fractions
\begin{equation}
f^+_L=\frac{|\bar A_0|^2}{\sum_\lambda |\bar A_\lambda|^2}\ ,\qquad
f^-_L=\frac{|A_0|^2}{\sum_\lambda |A_\lambda|^2}\,.
\end{equation}
We have therefore the relationships
\eq{
\begin{array}{ll}
\dis BR= \frac{1}{2} \frac{1}{\Gamma_{\rm total}} \left({\bar \Gamma}+ \Gamma \right)\ ,\quad
& \dis{\cal A}_{CP}=\frac{{\bar \Gamma} - \Gamma}{{\bar \Gamma} + \Gamma}\ , \\[4mm] 
\dis f_L = \frac{1}{2} \left( f_L^+ + f_L^- \right)\ , 
& \dis {\cal A}_{CP}^0=\frac{f_L^+ - f_L^-}{f_L^+ + f_L^-}\ .
\end{array}
}
and we can easily define the longitudinal observables in terms of these observables
\begin{eqnarray}
&&\hspace{-0.35cm} BR^{\rm long}\!=\!BR\cdot f_L \cdot [1+{\cal A}_{CP}^0 \cdot {\cal A}_{CP}]\,, \label{brL}\\
&&\hspace{-0.35cm} A^{\rm long}_{\rm dir} \!=\! -\frac{{\cal A}_{CP}^0 + {\cal A}_{CP}}{1+{\cal A}_{CP}^0 \cdot {\cal A}_{CP}}\,,\\
&&\hspace{-0.35cm} A^{\rm long}_{\rm mix} \!=\! \eta\sqrt{1-(A^{\rm long}_{\rm dir})^2}
 \sin(2\beta+{\rm arg}(A_0/{\bar A_0}))\,, \\
&&\hspace{-0.35cm} A^{\rm long}_{\rm \Delta \Gamma} \!=\! - \eta\sqrt{1-(A^{\rm long}_{\rm dir})^2}
 \cos(2\beta+{\rm arg}(A_0/{\bar A_0}))\,, \label{AdL} \hspace{1cm}
\end{eqnarray}
where the relative phase between $A_0$ and $\bar A_0$ is defined following the notation of Ref.~\cite{Aubert:2008zza}
\begin{equation} {\rm arg}(A_0/{\bar A_0})=2 \Delta \delta_0 + 2 \Delta \phi_0\,.
\end{equation}
In this analysis, the mixing angle $\phi_d$ was assumed to have its Standard Model value $2\beta=0.75\pm 0.03$ rad, which we used to translate the measured quantities into the longitudinal observables $A^{\rm long}_{\rm mix}$ and  $A^{\rm long}_{\rm \Delta \Gamma}$.
In the experimental analysis of Ref.~\cite{Aubert:2008zza}, there is only sensitivity to $ \sin(2\beta+{\rm arg}(A_0/{\bar A_0}))$, which means that there is a sign ambiguity in $\Adel^{\rm long}$, as seen in Eq.~(\ref{obslong}).

\section{Neutral-meson observables with a finite width difference}
\label{B}

Since we consider the $B_s$ meson in the present article, we have to discuss the modifications induced by a finite lifetime difference, before distinguishing its impact on flavour-tagged and flavour-untagged analyses for the longitudinal observables of interest. We will see that we must include different $O(\Delta \Gamma_s/\Gamma_s)$ corrections in each case to connect the measured quantities and the observables (\ref{eq:br})-(\ref{eq:amix}). As will become clear in the following, most of this discussion applies not only to two-body vector modes but more generally to any decay of a neutral meson into a $CP$ eigenstate.

\subsection{Branching ratios in presence of meson mixing}

The $B$-$\bar{B}$ systems can be described in terms of CP-conjugate flavour states $|B_Q\rangle$ and $|\bar{B}_Q\rangle$ ($Q=d,s$). The time evolution of an isolated neutral $B_Q$ meson of a given flavour at $t=0$ and decaying at a later time $t$ into a CP eigenstate $f$ is given as~\cite{Dunietz:1986vi,Dunietz:2000cr}
\begin{eqnarray}
&&\hspace{-0.9cm}
\Gamma(B_Q(t)\to f)= N_f \frac{|A_f|^2+|\bar{A}_f|^2}{2} e^{-\Gamma t} \times \Big[\cosh\frac{\Delta\Gamma t}{2} \nonumber \\
&& \hspace{-0.9cm}
+ A_{\rm dir} \cos(\Delta M t)-A_{\Delta\Gamma}
   \sinh\frac{\Delta\Gamma t}{2}+A_{\rm mix}\sin\Delta M t \Big]\,, \nonumber\\
&&\label{untagged}\\ 
&& \hspace{-0.9cm}
\Gamma(\bar{B}_Q(t)\to f)= N_f\frac{|A_f|^2+|\bar{A}_f|^2}{2}  e^{-\Gamma t}
 \times \Big[\cosh\frac{\Delta\Gamma t}{2} \nonumber\\
&&\hspace{-0.9cm}
- A_{\rm dir} \cos(\Delta M t)-A_{\Delta\Gamma}
   \sinh\frac{\Delta\Gamma t}{2}-A_{\rm mix}\sin\Delta M t\Big]\,,\nonumber\\
&&\label{untagged2}
\end{eqnarray}
where the Hamitonian eigenvalues $\mu_{L,H}=M_{L,H}-i\Gamma_{L,H}/2$ define $\Delta\mu=\mu_H-\mu_L=\Delta M-i\Delta\Gamma/2$. Also, $\Gamma\equiv (\Gamma_H+\Gamma_L)/2$ and $A_f=\langle f|B_Q\rangle$. $N_f$  is a time-independent, but state-dependent normalisation factor, corresponding to the integration over phase space. The mixing ratio $q/p=\exp(-i\phi_Q)$ is assumed to be a pure phase in the present article (as suggested by the very small values  of the flavour specific asymmetries both for $B_d$ and $B_s$ \cite{HFAG}). 

The normalisation factor $N_f$ comes from
\begin{equation}\label{eq:timeevolution}
\Gamma(B_Q(t)\to f)=N_f |\langle f| B_Q(t) \rangle|^2\,.
\end{equation}
Let us notice that this ``definition'' is rather ambiguous, as one generally considers initial and final states that are asymptotic (mass eigen)states, which $B_Q$ is not. However, one can still determine this factor by going back to the derivation of Fermi's golden rule. 
This is generally done for an initial mass eigenstate of the unperturbed Hamiltonian 
but can be easily adapted to the superposition of two mass eigenstates $|B_L\rangle, |B_H\rangle$, provided that the difference of energy between the two transitions $\omega_L-\omega_H=\langle f|H_1|B_L\rangle-\langle f|H_1|B_H\rangle$ is small. In that case, the normalisation $N_f$ corresponds to the phase space available, computed for the mass eigenstates of the unperturbed Hamiltonian. We obtain therefore the same normalisation as in the  case of charged $B$-decays
\begin{equation}
N_f=g_{ps} \Gamma\,,
\end{equation}
where $g_{ps}$ is given by the phase space with an incoming meson of mass $M=(M_H+M_L)/2$. 

Since there is no unambiguous definition of the CP-averaged branching ratio for neutral mesons as the states involved ($B_Q$ and $\bar{B}_Q$) are not mass eigenstates, we should define
what we call the CP-averaged branching ratio for $B_q$ decays. We opt for the simple definition, inspired by the charged meson case, and corresponding to the value that we would obtain through a measurement at $t=0$ (i.e., before neutral $B$-meson mixing could take place)
\begin{equation}\label{norm}
BR_{f}=g_{ps}(|A_f|^2+|\bar{A}_f|^2)/2\,.
\end{equation}
We will show later that this definition coincides exactly with 
time-integrated CP-averaged decay widths in the limit $\Delta\Gamma\to 0$, but that the relationship is corrected by terms of order $O(\Delta \Gamma/\Gamma)$. This correction depends on the experimental setting, because each one has a different sensitivity to the time evolution of the neutral $B$-meson. This correction factor can thus be seen as a correction for the temporal acceptance of the considered experiment.

Let us now come to $B_Q\to V_1V_2$ decays, and let us assume that all the direct asymmetries vanish, whereas $A_{0,\rm mix}=A_{||,\rm mix}=-A_{\perp,\rm mix}=\eta \sin\phi_Q$ and $A_{0,\Delta\Gamma}=A_{||,\Delta\Gamma}=-A_{\perp,\Delta\Gamma}=-\eta \cos\phi_Q$, where $\eta=\eta_{V_1}\eta_{V_2}$. 
As discussed in Ref.~\cite{Dunietz:2000cr}, we can obtain the time-dependent decay width from 
Eq.~(\ref{angularwidth}), upon the identification
\begin{eqnarray}
&&\hspace{-0.4cm} |a_f|^2 \to   N |A_f|^2  e^{-\Gamma t}\\
&&\hspace{-0.3cm}\times \Big[\cosh\frac{\Delta\Gamma t}{2}
+\eta_f\cos\phi_Q \sinh\frac{\Delta\Gamma t}{2} +\eta_f\sin\phi_Q\sin\Delta m t \Big]\,,\nonumber\\[3mm]
&&\hspace{-0.4cm}  {\rm Re}[a_0^* a_{||}]\to N |A_0||A_{||}|\cos(\delta_{||}-\delta_0) e^{-\Gamma t} \\
&&\hspace{-0.3cm}\times \Big[\cosh\frac{\Delta\Gamma t}{2}+\eta\cos\phi_Q\sinh\frac{\Delta\Gamma t}{2}
+\eta\sin\phi_Q\sin\Delta m t \Big]\,,\nonumber\\[3mm]
&&\hspace{-0.4cm} {\rm Im}[a_f^* a_\perp]\to N |A_f||A_\perp| e^{-\Gamma t} \times \Big[
\sin\delta_f \cos(\Delta m t)\label{imagpart}\\
&&\hspace{-0.3cm}-\eta\cos\delta_f\cos\phi_Q \sin(\Delta m t)+\eta\cos\delta_f\sin\phi_Q\sinh\frac{\Delta\Gamma t}{2}\Big]\,,\nonumber
\end{eqnarray}
with $f=0,\perp,\|$, and $\delta_0$ and $\delta_\|$ the relative strong phases of $A_0$ and $A_\|$ with respect to $A_\perp$. There is a common normalisation factor $N=g_{ps}\Gamma$ for all amplitudes. The CP-conjugate expression can be obtained by the replacement $\phi_Q\to-\phi_Q$ and by multiplying by $(-1)$ the two imaginary parts involving $a_\perp$ in Eq.~(\ref{imagpart}).

\subsection{Tagged analysis}

Flavour-tagged analyses are particularly easy to perform in $B$-factories as they produce intricated $B\bar{B}$ pairs~\cite{Aubert:2008zza}, and they allow one to study separately the time-dependence of $B$ and $\bar{B}$ samples as well as to extract the modulus and relative phases of the three transversity amplitudes through an angular analysis. 
The probability of the process depends on the decay times $t_{\rm tag}$ and $t_{CP}$ of both mesons (the one decaying into a tagging state $f_{\rm tag}$ and the one actually studied for CP-violation and decaying into $f_{CP}$). After integrating over the sum $t_{\rm tag}+t_{CP}$, one obtains for $\Gamma_{\rm tag}(B_Q(t)\to f)$ a structure similar to that for untagged decays provided that the exponential $\exp(-\Gamma t)$ is replaced by $\exp(-\Gamma |t|)$ and that $t=t_{CP}-t_{\rm tag}$ is allowed to run from $-\infty$ to $+\infty$. Indeed, from Ch.~1 in Ref.~\cite{Harrison:1998yr}, we have the joint decay amplitude $A(t_{\rm tag},t_{CP})$ which can be integrated over time to yield
\begin{eqnarray}
&&\hspace{-0.9cm} \int dt_{tag}\int dt_{CP} \int[dp] A(t_{\rm tag},t_{CP}) =\frac{\pi}{2}\frac{N_{{\rm tag}}}{\Gamma} |\bar{A}_{{\rm tag}}|^2\nonumber\\
&&\hspace{-0.9cm} \times \int_{-\infty}^\infty dt
N_{CP} \frac{|A_{CP}|^2+|\bar{A}_{CP}|^2}{2} e^{-\Gamma|t|}\times \Big[\cosh\frac{\Delta\Gamma t}{2}\nonumber\\
&&\hspace{-0.9cm} + A_{\rm dir} \cos(\Delta m t)
-A_{\Delta\Gamma} \sinh\frac{\Delta\Gamma t}{2}+A_{\rm mix}\sin\Delta m t \Big]  \label{tagged2}\\
&&\hspace{-0.9cm} = \frac{\pi}{2} BR(\bar{B}_Q\to f_{\rm tag})\times \int_{-\infty}^\infty dt\ \Gamma_{\rm tagged}(B_Q(t)\to f_{\sss CP})\nonumber\,.
   \end{eqnarray}
where the first factor comes from the angular integral describing the $e^+e^-\to B_Q\bar{B}_Q$ transition, and we have defined $\Gamma_{\rm tagged}(B_Q(t)\to f_{CP})$ as the integrand of Eq.~(\ref{tagged2}).

In the case of a non-vanishing width difference, and assuming the normalisation Eq.~(\ref{norm}), we can determine the CP-averaged branching ratio by considering
\begin{eqnarray}
\hspace{-1cm} BR_{f,{\rm tagged}}&=&\int_{-\infty}^{+\infty}\big[\Gamma_{\rm tagged}(B_Q(t)\to f)\nonumber\\
&&\hspace{-1.2cm} +\Gamma_{\rm tagged}(\bar{B}_Q(t)\to f)\big]  = BR_{f} \times \frac{\Gamma^2}{\Gamma_H\Gamma_L}\,.
\end{eqnarray}
We have therefore a correction of the time-integrated branching ratio with respect to the branching ratio in absence of mixing. This correction is due to the difference of the widths between the two neutral states.
This is typically a small correction: if we define $y\equiv \Delta\Gamma/(2\Gamma)$, we can write \cite{HFAG}
\eq{\hspace{-0.07cm}\frac{\Gamma^2}{\Gamma_H \Gamma_L}=\frac{1}{1-y^2}=\Bigg\{
\begin{array}{ll}
1+ (0\pm 2)\cdot 10^{-4} & {\rm for\ }B_d\,,\\
1+ (2\pm 2)\cdot 10^{-3} & {\rm for\ }B_s\ .
\end{array}
}

\subsection{Untagged analysis}

In the case of hadronic machines, such as CDF, D0 and LHCb, we encounter a rather different situation with no flavour tags available and an integral over time being performed. Since there is no information on the second $B$-meson being produced, one must consider Eqs.~(\ref{untagged}) and (\ref{untagged2}) for $t\geq 0$. The CP-averaged branching ratio can then be determined through
\begin{eqnarray}
\hspace{-0cm} BR_{f,{\rm untag}}&=&\int_0^{+\infty}\frac{1}{2}\big[\Gamma_{\rm untag}(B_Q(t)\to f)\\
&&\hspace{-2.1cm}+\Gamma_{\rm untag}(\bar{B}_Q(t)\to f)\big]= BR_{f} \times \frac{\Gamma^2}{\Gamma_L\Gamma_H}
  \left[1-A_{\Delta\Gamma}\frac{\Delta\Gamma}{2\Gamma}\right]\,.\nonumber
\end{eqnarray}
Compared to the tagged case, we see that there is a further term coming from $A_{f,\Delta\Gamma}$.
The resulting  correction for the branching ratio is larger than the one for B-factories, since it is linear in the small quantity $\Delta\Gamma/(2\Gamma)$, and not quadratic.

For vector-vector channels, assuming that the production rate for $B_Q$ and $\bar{B}_Q$ is the same and that there is no CP-violation in decay, $BR_{V_1V_2,{\rm untag}}$ yields an angular structure identical to Eq.~(\ref{angularwidth}), with the replacement
\begin{eqnarray} 
|a_f|^2 &\to& g_{ps} |A_{f}|^2\frac{\Gamma^2}{\Gamma_L\Gamma_H}\Bigg[1+\eta_f \frac{\Delta\Gamma}{2\Gamma}\cos\phi_Q\Bigg]\,,\qquad\\[3mm]
{\rm Re}[a^*_{0}a_{||}] &\to& g_{ps} |A_0||A_{||}| \cos(\delta_{||}-\delta_0)\nonumber\\
&&\times\frac{\Gamma^2}{\Gamma_L\Gamma_H}\Bigg[1+\eta\frac{\Delta\Gamma}{2\Gamma}\cos\phi_Q\Bigg]\,,\\[3mm]
{\rm Im}[a^*_{f}a_\perp] &\to & g_{ps} |A_f| |A_\perp|\nonumber\\
&&\times\frac{\Gamma^2}{\Gamma_L\Gamma_H}\Bigg[\eta_f \frac{\Delta\Gamma}{2\Gamma}\cos\delta_f\sin\phi_Q\Bigg]\,,
\end{eqnarray}
where the $\cos\phi$ terms are the remnants of $A_{\Delta\Gamma}$.
After integration over the angles, we obtain the CP-averaged width
\begin{eqnarray}
&&\Gamma_{B_Q\to V_1V_2}=
  g_{ps}\frac{\Gamma^3}{\Gamma_L\Gamma_H} \times \Bigg[|A_0|^2+|A_{||}|^2+|A_{\perp}|^2\nonumber \\
&&\qquad+\eta\frac{\Delta\Gamma}{2\Gamma}\cos\phi_Q (|A_0|^2+|A_{||}|^2-|A_{\perp}|^2)\Bigg]\,,
\end{eqnarray}
which is expressed in terms of the longitudinal and transverse CP-averaged branching ratios.

In the case of the LHCb analysis for $B_s\to K^{*0}\bar{K}^{*0}$ ($\eta=1$), an angular analysis was performed to extract the total branching ratio as well as the longitudinal polarisation. The expression in Ref.~\cite{LHCb} for the latter quantity was given in the SM case with a vanishing mixing angle $\phi_s$. However, the angular analysis performed does not rely on this assumption, so that we can just identify the coefficients with the same angular dependence in the differential decay width, leading to: $f_{L,LHCb}=|a_{0}^2|/\Gamma_{B\to V_1V_2}$.
We can thus derive the relation, valid in the absence of CP-violation in decay
\begin{equation}
f_{L,LHCb}\times BR_{LHCb}=BR^{\rm long}\frac{\Gamma_s^2}{\Gamma_L\Gamma_H}\left[1+\frac{\Delta\Gamma_s}{2\Gamma_s}\cos\phi_s\right]\,.
\end{equation}
We see that even in the SM case where $\phi_s$ is tiny, there is a small contribution from $\Delta\Gamma_s/\Gamma_s$ to the relationship between the polarisation measured at LHCb and $BR^{\rm long}$.


\begin{thebibliography}{99}

 \bibitem{CKM}
  A.~Lenz, U.~Nierste, J.~Charles, S.~Descotes-Genon, A.~Jantsch, C.~Kaufhold, H.~Lacker, S.~Monteil {\it et al.},
  Phys.\ Rev.\  {\bf D83}, 036004 (2011).
  
  
\bibitem{Charles:2011va}
  J.~Charles, O.~Deschamps, S.~Descotes-Genon, R.~Itoh, H.~Lacker, A.~Menzel, S.~Monteil, V.~Niess {\it et al.},
  Phys.\ Rev.\  {\bf D84 } (2011)  033005.

\bibitem{lunghisoni}
  E.~Lunghi, A.~Soni,
  Phys.\ Lett.\  {\bf B697}, 323-328 (2011).
  
\bibitem{gino}
  R.~Barbieri, P.~Campli, G.~Isidori, F.~Sala, D.~M.~Straub,
  [arXiv:1108.5125 [hep-ph]].
  
\bibitem{CDF}
CDF Collaboration, CDF Note 10206, November 2, 2010, http://www-cdf.fnal.gov/ 
physics/new/bottom/100513. blessed-BsJpsiPhi 5.2fb/cdf10206 sin2betas.pdf 

\bibitem{D0}
S. Burdin [D0 Collaboration], ``Measurements of CP violation in the Bs system at D0'', talk given at the Europhysics Conference on High-Energy Physics 2011, July 21 2011.
http://indico.in2p3.fr/getFile.py/access?con tribId=1005\&sessionId=2\&resId=0\&materialId=slides\& confId=5116 

\bibitem{LHCb}
LHCb Collaboration, LHCb-CONF-2011-049. 

\bibitem{D01}
  V.~M.~Abazov {\it et al.} [D0 Collaboration],
  Phys.\ Rev.\  {\bf D82}, 032001 (2010).
  
\bibitem{D02}
  V.~M.~Abazov {\it et al.} [D0 Collaboration],
  Phys.\ Rev.\ Lett.\  {\bf 105}, 081801 (2010).

\bibitem{apo}
  S.~Descotes-Genon, J.~Matias and J.~Virto,
  Phys.\ Rev.\  D {\bf 76} (2007) 074005.
  
\bibitem{prl}
  S.~Descotes-Genon, J.~Matias and J.~Virto,
  Phys.\ Rev.\ Lett.\  {\bf 97}, 061801 (2006).
  
 \bibitem{adeva}
R.~Aaij {\it et al.} [LHCb Collaboration], arXiv:1111.4183.

\bibitem{Aubert:2008zza}
  B.~Aubert {\it et al.}  [BABAR Collaboration],
  Phys.\ Rev.\  D {\bf 78}, 092008 (2008).

  
\bibitem{belle}
  K.~F.~Chen {\it et al.}  [BELLE Collaboration],
  Phys.\ Rev.\ Lett.\  {\bf 94}, 221804 (2005).
  
\bibitem{pro}
  J.~Virto,
  AIP Conf.\ Proc.\  {\bf 964}, 90-94 (2007).
  
\bibitem{QCDF}
M.~Beneke and M.~Neubert,
Nucl.Phys.B {\bf 675} (2003) 333.

\bibitem{BVV}
  M.~Beneke, J.~Rohrer and D.~Yang,
  Nucl.\ Phys.\  {\bf B774}, 64-101 (2007).
 
 \bibitem{kagan}
  A.~L.~Kagan,
  Phys.\ Lett.\  {\bf B601}, 151-163 (2004).
  
\bibitem{HFAG}
Heavy Flavor Averaging Group (HFAG),  http://www. slac.stanford.edu/xorg/hfag/.

\bibitem{LHCbNote}
LHCb Collaboration, note LHCb-CONF-2011-019.

\bibitem{Dunietz:2000cr}
  I.~Dunietz, R.~Fleischer and U.~Nierste,
  Phys.\ Rev.\  D {\bf 63} (2001) 114015.
  
\bibitem{Dunietz:1986vi}
  I.~Dunietz, J.~L.~Rosner,
  Phys.\ Rev.\  D {\bf 34} (1986) 1404.
  
\bibitem{Harrison:1998yr}
  P.~F.~Harrison and H.~R.~Quinn Eds.
  ``The BABAR physics book: Physics at an asymmetric B-factory''.

\end{thebibliography}
\end{document}